\documentclass[10pt]{article}
\usepackage{fancyhdr}
\usepackage{extramarks}
\usepackage{amsmath}
\usepackage{amsthm}
\usepackage{amsfonts}
\usepackage{siunitx}
\usepackage{tikz}
\usepackage[plain]{algorithm}
\usepackage{algpseudocode}
\usepackage{multirow}
\usepackage{booktabs}
\usepackage{graphicx}
\usepackage{subfigure}
\usepackage[colorlinks,linkcolor=black,anchorcolor=black,citecolor=black,urlcolor=blue]{hyperref}
\usepackage{amsmath,bm}
\usepackage{booktabs}
\usepackage{mathtools}
\usepackage{amssymb}
\usepackage{caption}
\usepackage{capt-of}
\usepackage{mciteplus}
\usepackage{cite}
\usepackage{mathrsfs}
\usepackage[title,titletoc,toc]{appendix}
\usepackage{xr}
\usepackage{parskip}
\usepackage{soul}
\usepackage{textcomp}
\usepackage[colaction]{multicol}
\usepackage[switch]{lineno}
\usepackage{lipsum}
\usepackage{etoolbox}
\usepackage{longtable}
\usepackage{array}
\usepackage{tablefootnote}
\usepackage{ragged2e}
\newcolumntype{C}[1]{>{\centering\arraybackslash}p{#1}}
\usetikzlibrary{automata,positioning}
\usetikzlibrary{automata,positioning}

\newtheorem{Def}{Defination}[section]

\begin{document}

\title{Persistent Laplacian projected Omicron BA.4 and BA.5 to become new dominating  variants}

\author{ Jiahui Chen$^1$, Yuchi Qiu$^1$, Rui Wang$^1$, and Guo-Wei Wei$^{1,2,3}$\footnote{
 		Corresponding author.		Email: weig@msu.edu} \\
 $^1$ Department of Mathematics, \\
 Michigan State University, MI 48824, USA.\\
East Lansing, MI 48823 USA.\\
 $^2$ Department of Electrical and Computer Engineering,\\
 Michigan State University, MI 48824, USA. \\
 $^3$ Department of Biochemistry and Molecular Biology,\\
 Michigan State University, MI 48824, USA. \\
 }
\date{\today} 

\maketitle

\begin{abstract}
Due to its high transmissibility, Omicron  BA.1 ousted the Delta variant to become a dominating variant in late 2021 and was replaced by more transmissible Omicron BA.2 in March 2022. An important question is  which new variants will dominate in the future. Topology-based deep learning models have had tremendous success in  forecasting emerging variants in the past. However, topology is insensitive to homotopic shape variations in virus-human protein-protein binding, which are crucial to viral evolution and transmission. This challenge is tackled with persistent Laplacian,  which is able to capture both the topology and shape of data. Persistent Laplacian-based deep learning models are developed to systematically evaluate variant infectivity. Our comparative analysis of  Alpha, Beta, Gamma, Delta, Lambda, Mu, and Omicron BA.1, BA.1.1, BA.2,  BA.2.11,  BA.2.12.1, BA.3,  BA.4, and BA.5 unveils that Omicron  BA.2.11,  BA.2.12.1, BA.3,  BA.4, and BA.5 are more contagious than BA.2. In particular, BA.4  and BA.5 are about 36\% more infectious than BA.2 and are projected to become new dominating variants by natural selection. Moreover, the proposed models outperform the state-of-the-art methods on three major benchmark datasets for mutation-induced protein-protein binding free energy changes.
\end{abstract}
\pagebreak
%
 {\setcounter{tocdepth}{4} \tableofcontents}
 \newpage
 %

\setcounter{page}{1}
\renewcommand{\thepage}{{\arabic{page}}}


%
 
\section{Introduction}
The coronavirus disease, 2019 (COVID-19) caused by severe acute respiratory syndrome coronavirus 2 (SARS-CoV-2) has lasted for more than years. The development of effective vaccines, monoclonal antibodies (mABs), and antiviral drugs has significantly improved our ability to bring COVID-19 pandemic under control. Nonetheless, the emerging SARS-CoV-2 variants become a major threat to existing vaccines, monoclonal antibodies (mABs), and antiviral drugs. 

The Omicron variant has   mutations  on various SARS-CoV-2 proteins, such as non-structure protein 3 (NSP3), NSP4, NSP5, NSP6, NSP12, NSP14, spike (S) protein, envelope protein, membrane protein, and nucleocapsid protein. Specifically, Omicron has three main lineages, BA.1 (B.1.1.529.1), BA.2 (B.1.1.529.2), and BA.3 (B.1.1.529.3), and many sub-lineages. Many new recombinants occurred, including  XD, XE, and XF. XD and XE are recombination of Delta and BA.1, while XE is basically a BA.2 Omicron lineage carrying a piece of BA.1 at the front end of its genome. The S protein of XE is still BA.2.

The research community focuses its attention on the mutations at the S protein receptor-binding domain (RBD)  due to the fact that the RBD facilitates the binding between the S protein and the host angiotensin-converting enzyme 2 (ACE2), which initiates the viral entry of a host cell and infection. It turns out that the binding strength  between the S protein RBD and the ACE2 is proportional to the viral infectivity \cite{li2005bats,qu2005identification,song2005cross,hoffmann2020sars,walls2020structure}. An artificial intelligence (AI) study revealed that natural selection is the governing mechanism for SARS-CoV-2 evolution \cite{chen2020mutations}. Specifically, viral evolution selects those mutations that are able to strengthen the RBD-ACE2 binding. This mechanism led to the occurrence of many variants, such as Alpha, Beta, Gamma, Delta, Mu, etc.  Natural selection in SARS-CoV-2 mutations was conformed beyond doubt in April 2021 by the genotyping of over half a million viral genomes isolated from patients \cite{wang2021vaccine}. 

Additionally,  antibodies are generated by the human immune response to infection or vaccination. A strong RBD-antibody  binding would  lock off RBD-ACE2 binding and directly neutralize the virus \cite{wang2020human,yu2020receptor,li2021impact}. As such,  mABs targeting the S protein, particularly the RBD, are designed to treat viral infection. It was unveiled that  viral evolution also selects those mutations that are able to weaken RBD-antibody  binding, leading vaccine breakthrough infections  \cite{wang2021mechanisms,wang2022emerging}. Therefore, a new virus with RBD mutations that make the virus more infectious and more capable of evading the antibody protection would become the next dominating variant, which is the underlying principle for the successful forecasting of Omicron BA.2's dominance \cite{chen2022omicron2}.  
 
In biophysics,  the strength of protein-protein complex is measured by binding free energy (BFE). Mutation-induced BFE change $\Delta\Delta G$ is calculated by
\begin{equation}
    \Delta\Delta G = \Delta G_\text{WT} - \Delta G_\text{MT}
\end{equation}
where $\Delta G_\text{WT}$ and $\Delta G_\text{MT}$ are the BFE of wild type and mutant. A positive (negative) BFE change indicates the strengthening 
(weakening) of the protein-protein binding. Protein-protein BFE changes can be carried out in a variety of ways as shown in software packages FOLDX \cite{guerois2002predicting}, SAAMBE \cite{petukh2016saambe},  mCSM-AB \cite{pires2016mcsm}, mCSM-PPI2 \cite{rodrigues2019mcsm}, BindProfX \cite{xiong2017bindprofx}, etc.  AI approaches take the advantage of existing data and often outperform other methods when experimental data become available.  Due to the structural complexity and high dimensionality of of protein-protein interactions (PPIs), methods that are able to effectively reduce the PPI structural complexity and dimensionality have demonstrated great advantages in predicting PPI BFE changes \cite{wang2020topology}. Advanced mathematics, particularly, persistent homology \cite{frosini1992measuring,edelsbrunner2000topological,zomorodian2005computing,carlsson2009topology, mischaikow2013morse,KLXia:2014c}, offers tremendous abstraction of PPIs. Persistent homology is the main workhorse in popular topological data analysis (TDA) \cite{de2007coverage, YaoY:2009,bubenik2014categorification,dey2014computing}. Element-specific persistent homology (EPH) has had tremendous success  in computational biology \cite{cang2017topologynet,cang2018representability} and  worldwide competitions in computer-aided drug design \cite{nguyen2019mathematical}. 

Based on FPH, a topology-based network tree (TopNetTree) model was constructed from conventional neural network and decision trees for predicting PPI BFE changes \cite{wang2020topology}. In the past two years, this approach has been  extended with SARS-CoV-2 related deep mutational data to predict the BFE changes RBD-ACE2 and RBD-antibody complexes up on RBD mutations \cite{chen2021prediction,chen2021revealing}. Initially, in early 2020, TopNetTree model was applied to successfully predict that RBD residues 452 and 501 ``have high chances to mutate into significantly more infectious COVID-19 strains'' \cite{chen2020mutations}. These RBD mutations later appeared in all major variants, Alpha, Beta, Delta, Gamma, Delta, Epsilon, Theta, Kappa, Lambada, Mu, and Omicron L452R/Q and N501Y mutations. In April 2021, the TopNetTree model predicted  a list of 31 RBD antibody-escape mutations, including W353R, I401N, Y449D, Y449S, P491R, P491L, Q493P, etc.  \cite{wang2021vaccine}. Notably, experimental results confirmed that mutations at RBD residues Y449, E484, Q493, S494, and Y505 enable the virus to escape antibodies \cite{alenquer2021signatures}. It was revealed that variants found in the United Kingdom and South Africa in late 2020 would strengthen virus infectivity, which is consistent with the experimental results \cite{dupont2021neutralizing}. In summer 2021, a topology-based deep neural network trained with mAbs (TopNetmAb) was developed to forecast a list of most likely vaccine-escape RBD mutations, such as S494P, Q493L, K417N, F490S, F486L, R403K, E484K, L452R, K417T, F490L, E484Q, and A475S \cite{chen2021revealing}, and mutations   S494P, K417N, E484K/Q, and L452R were designated as the variants of concern or variants of interest denounced by the Worldwide Health Organization (WHO).  The correlation between the experimental deep mutational data \cite{linsky2020novo} and  AI-predicted RBD-mutation-induced BFE changes for all possible 3686 RBD mutations on the RBD-ACE2 complex is 0.7 \cite{chen2021revealing}. In comparison,  experimental deep mutational results for the same set of RBD mutations from 2 different labs  only have a correlation of 0.67 \cite{starr2020deep,linsky2020novo}. TopNetmAb predictions of Omicron \cite{chen2022omicron} and Omicron BA.2 	\cite{chen2022omicron2} infectivity, vaccine breakthrough, and antibody resistance were nearly perfectly confirmed by experiments and pandemic evolution in the world. These mechanistic discovery and successful predictions  may not be achievable via purely experimental means, indicate the indispensable role of AI for scientific discovery.

However, persistent homology and TDA provide only topological invariants, which may not be sufficient for representing PPI data. In particular, the shape of data arisen from a family of homotopy geometries cannot be captured by persistent homology. For example, the geometry of each drum in an acoustic drum set is designed to offer a specific sound or frequency, but persistent homology is insensitive to the change in the sizes (or shapes) in the drum set. This challenge in TDA was addressed by the introduction of persistent Laplacian, or persistent spectral graph \cite{wang2020persistent}. Persistent Laplacian manifests the full set of topological invariants and the shape of data in its harmonic and non-harmonic spectra, respectively. Additional mathematical analysis \cite{memoli2020persistent} and a software package, i.e., HERMES \cite{wang2021hermes}, for persistent Laplacian have been reported in the literature.  This method has been successfully applied to biological studies,  including protein thermal stability\cite{wang2020persistent}, protein-ligand binding  \cite{meng2021persistent}, and protein-protein binding problems \cite{wee2022persistent}.

In the present work, we introduce element-specific and site-specific persistent Laplacians to forecast emerging SARS-CoV-2 variants. We hypothesize that persistent Laplacians  generates intrinsically low-dimensional representations of PPIs and dramatically reduce the dimensionality of PPI data, leading to a reliable high-throughput screening of emerging SARS-CoV-2 variants. To quantitatively validate this hypothesis, we integrate the harmonic and non-harmonic spectra of persistent Laplacians with efficient machine learning algorithms, i.e., gradient boosting tree (GBT) and deep neural network (Net), to predict PPI $\Delta\Delta G$ following mutations. The resulting topological and spectral-based machine learning models are validated on three major benchmark datasets, the AB-Bind database  \cite{sirin2016ab}, SKEMPI dataset \cite{moal2012skempi} and SKEMPI v2.0 dataset \cite{jankauskaite2019skempi}, giving rise to the state of the art performance. Meanwhile, with additional training on SARS-CoV-2 related datasets, our models forecast  emerging SARS-CoV-2 variants and  recommend  four Omicron subvariants, i.e., BA.2.11,  BA.212.1, BA.4, and BA.5 for active  surveillance.

\section{Results}
In this section, we first carry out the infectivity predictions on emerging SARS-CoV-2 variants. Next,  three benchmark PPI datasets, i.e., the AB-Bind  \cite{sirin2016ab},  SKEMPI  \cite{moal2012skempi}, and   SKEMPI 2.0 datasets \cite{jankauskaite2019skempi}  are employed to demonstrate the proposed  persistent Laplacian-based AI models  with  ten-fold cross validations. Two evaluation metrics,  Pearson correlation $R_p$ and the root-mean-square error (RMSE), are used to assess the quality of the present models. Lastly, we present the validation  of our models on SARS-CoV-2-related datasets.

\subsection{Emerging SARS-CoV-2 variants: Infectivity}

\begin{figure}[h]
	\centering
	\includegraphics[width = 0.95\textwidth]{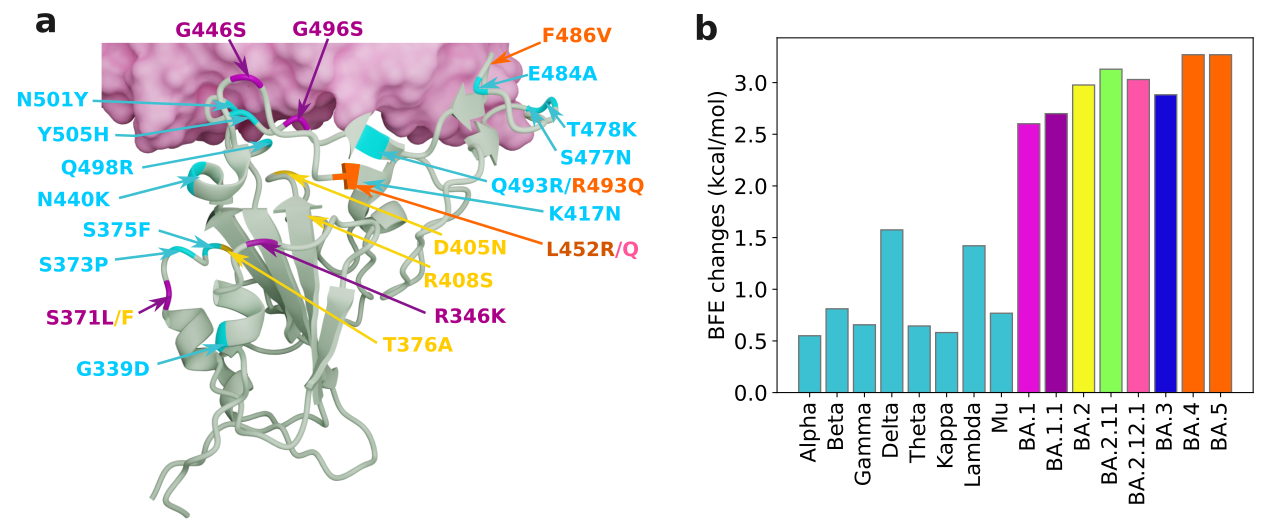}
	\caption{ 
	The RBD mutations of Omicron subvariants at the RBD-ACE2 interface and their mutation-induced BFE changes. 
{\bf a} RBD mutations of Omicron subvariants at the RBD-ACE2 interface (PDB: 7T9L \cite{mannar2021sars}). 
The shared 12 mutations are shown in cyan. 
BA.1  mutations are plotted with magenta. 
BA.2  mutations are marked in yellow. 
BA.4 and BA.5  mutations are labeled in orange. 
The rest colors can be matched from the right chart. 
{\bf b} A comparison of predicted mutation-induced BFE changes for various SARS-CoV-2 variants and subvariants.
	}
	\label{fig:omicron}
\end{figure}

Figure~\ref{fig:omicron} shows the RBD mutations of Omicron subvariants and their  BFE changes of SARS-CoV-2 variants.
A comparison is also given to other main SARS-CoV-2 variants Alpha, Beta, Gamma, Delta, Theta, Kappa, Lambda, and Mu variants. The Delta variant had the highest BFE change among the earlier variants and was the most infectious variant before the occurrence of the Omicron variant, which explains its dominance in 2021. 
Omicron BA.1, BA.2, and BA.3 have the common RBD mutations G339D, S373P, S375F, K417N, N440K, S477N, T478K, E484A, Q493R, Q498R, N501Y, and Y505H.
Omicron BA.1 has three distinct RBD mutations S371L, G446S, and G496S. Four distinct mutations, S371F, T376A, D405N, and R408S, were found for Omicron BA.2. Omicron BA.3 shares three mutations either with BA.1 or BA.2: S371F, D405N, and G446S. 
The AI-predicted BFE changes of BA.1, BA.2, and BA.3 are 2.60, 2.98, and 2.88 kcal/mol, respectively \cite{chen2022omicron2}. These values are significantly higher than those of other major SARS-CoV-2 variants as shown in Figure~\ref{fig:omicron}. Note that Omicron BA.2 is the most infectious variant. It is about 20 and 4.2 times as infectious as the original SARS-CoV-2 and the Delta variant, respectively. The machine learning model also predicts that BA.2 is about 1.5 times as contagious as BA.1, which is highly consistent with experimental studies \cite{timesofisrael,lyngse2022transmission}. BA.2 has been the dominating variant since late March 2022 \cite{chen2022omicron2}.  

We have also examined the other Omicron subvariants, namely, BA.1.1,   BA.2.11,  BA.2.12.1, BA.4, and BA.5. Compared with BA.1, BA.1.1 has one additional RBD mutation, i.e.,  R346K.  BA.2.11 has one more RBD mutation, L452R, than BA.2 does.  BA.2.12.1 has an extra RBD mutation, L452Q, compared with BA.2.  BA.4 and BA.5 share the same set of RBD mutations but differ in ORF7b, nucleocapsid (N), and membrane (M) proteins. 
They have three additional RBD mutations, L452R, F486V, and R493Q compared with BA.2. Note that  R493Q  is a reversion to the wide type, Q493. It is interesting that L452R is one of Delta's two RBD mutations. Additionally, mutations simultaneously occurred on two  RBD residues, L452 and N501, which were singled out by our AI model in early 2020 \cite{chen2020mutations}.    

Our AI-predicted BFE changes for BA.1.1,  BA.2.11,  BA.2.12.1, BA.4, and BA.5   are 2.70, 3.13, 3.03, 3.27, and 3.27  kcal/mol, respectively. It is noticed that BA.4 and BA.5 
are predicted to be 1.36 times as infectious as BA.2 and have high potential to become new dominating SARS-CoV-2 variants. 

\subsection{The performance on the AB-Bind dataset}

The AB-Bind dataset,  including 1,101 mutational data entries for experimentally determined BFE changes  \cite{sirin2016ab} is considered in the validation of the proposed models. Its 645 single mutations involving 29 antibody-antigen complexes are denoted as the AB-Bind S645 set. In the AB-Bind S645 set, about one-fifth of mutations strengthen the binding, while the rest are destabilizing mutations. In particular, 27 non-binders, which are mutants determined not to bind within the experimental sensitivity of the assay, are in the dataset. The mutation-induced binding free energy changes for these non-binders were set to -8 kcal/mol. For machine learning models, non-binders are outliers and can cause a very negative impact on model accuracy.

\begin{table}[htb]
	\centering
	\begin{tabular}{ll||ll}
		\toprule
		Method & $R_p$ & Method & $R_p$ \\
		\midrule
		TopLapGBT  & 0.89/$0.95^*$ & mCSM-AB & 0.53/$0.56^*$\\
		LapGBT & 0.88/$0.94^*$ & Discovery Studio & 0.45\\
		TopGBT  & 0.88/$0.95^*$ & mCSM-PPI & 0.31 \\
		TopLapNetGBT  & 0.87/$0.93^*$ & FoldX & 0.34 \\
		LapNetGBT  & 0.87/$0.91^*$ & STATIUM & 0.32 \\
		TopNetGBT  & 0.86/$0.93^*$ & DFIRE & 0.31 \\
		TopNet  & 0.81/$0.88^*$ & bASA & 0.22 \\
		TopLapNet  & 0.79/$0.87^*$ & dDFIRE & 0.19\\
		LapNet  & 0.72/$0.81^*$ & Rosetta & 0.16\\
		\midrule
	\end{tabular}
	\vspace{3mm}
	\caption[caption]{
		Comparison of the Pearson correlation coefficients ($R_p$) of various methods for the AB-bind  S645  set. Except for present TopLapGBT and TopLapNet, the results of other existing methods are adopted from Ref.  \cite{pires2016mcsm}. \\\hspace{\textwidth}
		$^*$Results exclude  27 non-binders (their $\Delta\Delta$Gs were set to -8 kcal/mol \cite{sirin2016ab}). } 
	\label{tab:ModelPerformance}
\end{table}

As shown in Table \ref{tab:ModelPerformance}, our TopLapGBT and LapNet  models  achieved the $R_p$ of 0.89 and 0.72 for the AB-Bind S645 set.  
In comparison,   TopNet  outperforms LapNet because TopNet includes  auxiliary features, while LapNet has only Laplacian features. 
The $R_p$ values of  our other seven models are lower than 0.89 but higher than 0.72. Note that our worst model (LapNet) still outperforms the other best model in the literature by a large margin of 36\%, while our best model is about 68\% better than the other best model in the literature, indicating the predictive power of our topology and Laplacian-based machine learning models. Both GBTs and Nets models are quite sensitive to system errors as the model training is based on optimizing the mean-square error of the loss function. The BFE changes of 27 non-binders (-8 kcal/mol) did not follow the distribution of the whole dataset. For the TopLapGBT model, the RMSE of AB-Bind S645set is 1.68 kcal/mol and reduces to 0.97 kcal/mol when 27 non-binder samples are excluded. In this case,  the $R_p$ of the TopLapGBT model is increased from 0.89 to 0.95. 
The consensus results of GBT and Net have correlations of 0.86-0.87, which are lower than that of GBT but higher than that of Net. GBT models outperform Net models in the validation, showing that GBT performs better than Net on a small dataset. 

\subsection{The performance on the SKEMPI dataset}
The SKEMPI dataset \cite{moal2012skempi} has 3,047 entries of BFE changes induced by mutations. This dataset is collected from the literature for protein-protein heterodimeric complexes with experimentally determined structures. It consists of single- and multi-mutations. Among them, 2,317 single mutations out of 3,047 entries are called the S2317 dataset. Recently, a subset of 1,131 non-redundant interface single-mutations is selected and denoted as the S1131 set \cite{xiong2017bindprofx}. Table~\ref{tab:skempi} shows the Pearson correlation coefficients on tenfold cross-validations of various models, including topology- and Laplacian-based models. The proposed topology- and Laplacian-based models are found to be more accurate than other existing methods. One may notice that for a larger training set, the consensus predictions of GBT and Net outperform GBT methods. Additionally, topology-based models contain topology features and auxiliary features, which include more biomolecular information than Laplacian-based models.

\begin{table}[ht!]
	\centering
	\begin{tabular}{ll||ll}
		\toprule
		Method & $R_p$ & Method & $R_p$ \\
		\midrule
		TopLapNetGBT & 0.87 & BindProfX & 0.738\\
		TopNetGBT & 0.87 & Profile-score+FoldX  & 0.738\\
		TopLapNet & 0.86 & Profile-score & 0.675\\
		TopNet & 0.86 & SAAMBE & 0.624 \\
		TopLapGBT & 0.86 & FoldX & 0.457\\
		TopGBT & 0.86 & BeAtMuSic &0.272\\
		LapNetGBT & 0.81 & Dcomplex &0.056 \\
		LapNet & 0.81 & & \\
		LapGBT & 0.78 & & \\
		\midrule
	\end{tabular}
	\vspace{3mm}
	\caption[caption]{
		Comparison of the Pearson correlation coefficients ($R_p$) of various methods for the S1131 set in the SKEMPI dataset. The results of other methods are adopted  from Ref. \cite{xiong2017bindprofx}. \\
		\hspace{\textwidth}
	} 
	\label{tab:skempi}
\end{table}

\subsection{The performance on the SKEMPI 2.0 dataset}

The SKEMPI 2.0 \cite{jankauskaite2019skempi} database is an updated version of the original SKEMPI database with new mutations from three other databases: AB-bind \cite{sirin2016ab}, PROXiMATE \cite{jemimah2017proximate}, and dbMPIKT\cite{liu2018dbmpikt}. This dataset has 7,085 entries, including single-mutations and multi-mutations. To validate mCSM-PPI2, David et al. filtered only single-point mutations, selected 4169 variants in 319 different complexes, and denoted them as the S4169 set  \cite{rodrigues2019mcsm}. Additionally,  set S8338 was derived from set S4169 by setting the BFE changes of the reverse mutations as  the negative values of the original BFE changes induced by mutations. We present our tenfold cross-validation results on sets S4169 and S8338 in Table~\ref{tab:skempi2}.   For S4169, TopLapNetGBT has the most accurate result with $R_p$ of 0.82 and RMSE of 1.06 kcal/mol. Topology-based models, aided by auxiliary features, have correlations greater than 0.80 and RMSE from 1.04 kcal/mol to 1.10 kcal/mol. 
Purely Laplacian-based models also performed quite well, with the Pearson correlation of 0.76, which is the same as that of the mCSM-PPI2.
 
\begin{table}[htb]
	\centering
	\begin{tabular}{ll||ll}
		\toprule
		\multicolumn{2}{c||}{S4169} & \multicolumn{2}{c}{S8338} \\\hline
		Method & $R_p$ & Method & $R_p$ \\
		\midrule
		TopLapNetGBT & 0.82 & TopLapNetGBT & 0.87\\
		TopNetGBT & 0.82 & TopLapNet & 0.87\\
		TopLapNet & 0.81 & TopNetGBT & 0.87\\
		TopLapGBT & 0.81 & TopNet & 0.86 \\
		TopNet & 0.81 & TopLapGBT & 0.85 \\
		TopGBT & 0.80 & TopGBT & 0.85 \\
		LapNetGBT & 0.77 & LapNetGBT &0.83 \\
		mCSM-PPI2 & 0.76 & mCSM-PPI2 & 0.82\\
		LapNet & 0.76 & LapNet & 0.81 \\
		LapGBT & 0.76 & LapGBT & 0.80 \\
		\midrule
	\end{tabular}
	\vspace{3mm}
	\caption[caption]{
		Comparison of the Pearson correlation coefficients ($R_p$) of various methods for  S4169 set and S8338 set in  SKEMPI 2.0. Results of mCSM-PPI2 are from Ref.\cite{rodrigues2019mcsm}
	} 
	\label{tab:skempi2}
\end{table}

For the S8338 set, TopLapNetGBT  has the highest Pearson correlation $R_p$ of 0.8702 and RMSE of 1.01 kcal/mol as shown in Table \ref{tab:skempi2}. TopLapNet has the most accurate results with $R_p$ of 0.8688 and RMSE of 0.984 kcal/mol. Topology models, aided by auxiliary features, have the $R_p$ in the range of (0.848, 0.870) and RMSE in the range of (1.070 kcal/mol, 0.984 kcal/mol). LapNet and LapGBT models have their $R_p$ values slightly lower than that of mCSM-PPI2, but the $R_p$ of their consensus (LapNetGBT) is higher than that of  the  mCSM-PPI2.

\subsection{The performance on SARS-CoV-2 datasets}

Training datasets have the utmost importance in implementing our machine learning model for SARS-CoV-2 applications. First, all the datasets mentioned above, including  AB-bind,\cite{sirin2016ab} PROXiMATE\cite{jemimah2017proximate}, dbMPIKT\cite{liu2018dbmpikt}, SKEMPI\cite{moal2012skempi}, and SKEMPI 2.0\cite{jankauskaite2019skempi}, are used in our model training. Additionally,  SARS-CoV-2-related datasets are also employed  to improve the prediction accuracy after a label transformation. These are  deep mutational enrichment ratio data, including  mutational scNeting data of ACE2 binding to the receptor-binding domain (RBD) of the S protein \cite{procko2020sequence}, mutational scNeting data of RBD binding to ACE2 \cite{starr2020deep, linsky2020novo}, and mutational scNeting data of RBD binding to CTC-445.2 and of CTC-445.2 binding to the RBD \cite{linsky2020novo}. Note that in our validation, our training datasets exclude the test dataset, which is a mutational scNeting data of RBD binding to ACE2. Here, these datasets provide more information on SARS-CoV-2 and can be used to calibrate the models to predict the real experimental results. 

\begin{figure}[h]
	\centering
	\includegraphics[width = 0.95\textwidth]{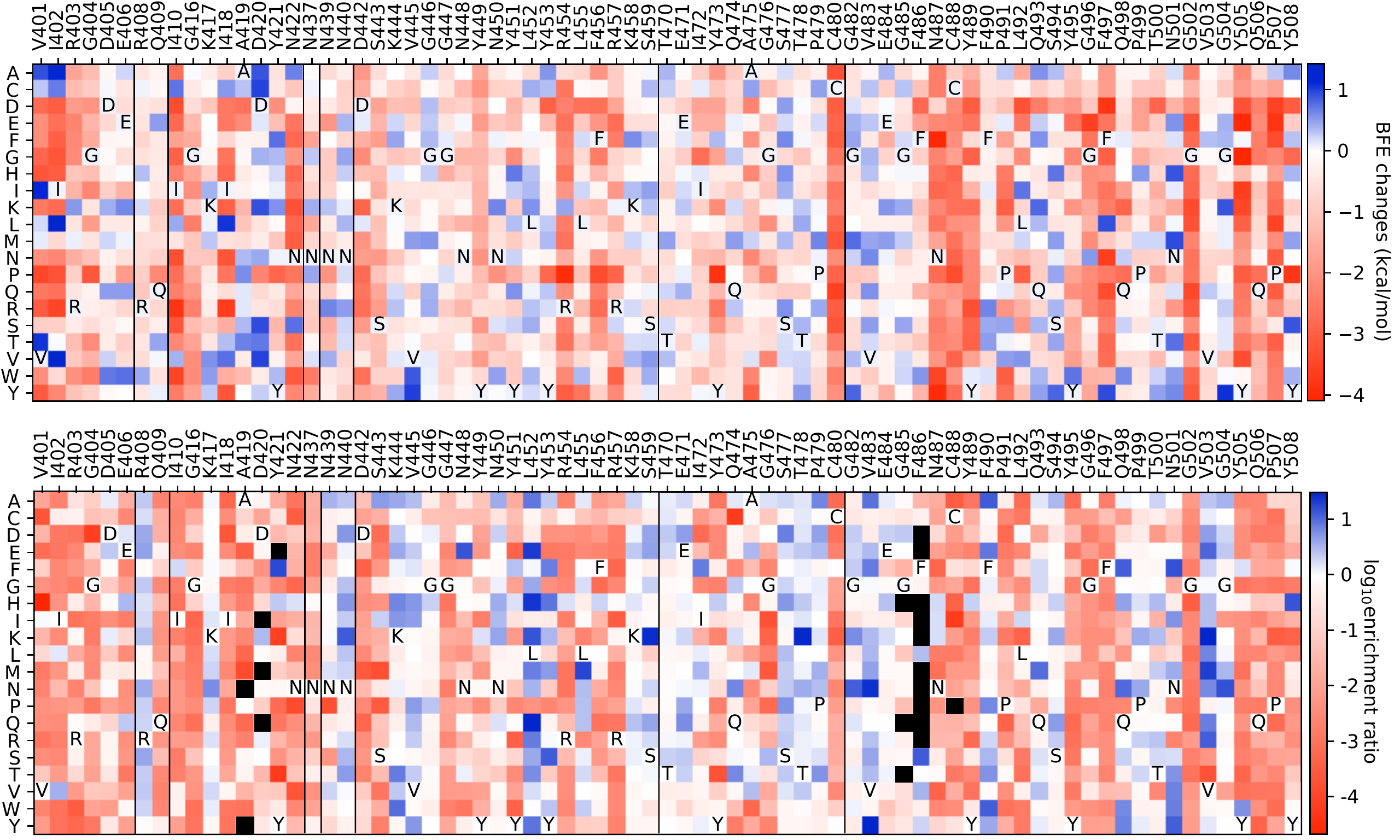}
	\caption{A comparison between experimental RBD deep mutation enrichment data and predicted BFE changes for SARS-CoV-2 RBD binding to ACE2 (6M0J) \cite{linsky2020novo}.  {\bf Top}: machine learning predicted BFE changes for single-site mutants of the S protein RBD. {\bf Bottom}: deep mutational scanning heatmap showing the average effect on the enrichment for single-site mutants of RBD when assayed by yeast display for binding to the S protein RBD \cite{linsky2020novo}.}
	\label{fig:validation}
\end{figure}

Here, we present a validation of our model BFE change prediction for mutations on S protein RBD compared to the experimental deep mutational enrichment data \cite{linsky2020novo}. We compare between experimental deep mutational enrichment data and BFE change predictions on SARS-CoV-2 RBD binding to ACE2 in Figure~\ref{fig:validation}. Both BFE changes (Figure~\ref{fig:validation} top) and enrichment ratios (Figure~\ref{fig:validation} bottom) describe the binding affinity changes of the S protein RBD-ACE2 complex induced by mutations.
It can be found that the predicted BFE changes are highly correlated to the enrichment ratio data. Pearson correlation is 0.69.

\section{Theories and methods}
This section presents brief reviews of spectral graph theory, simplicial complex, and persistent Laplacian are presented. Machine learning  and deep learning models are discussed in test datasets and validation settings.

\subsection{Persistent Laplacians} 

\subsubsection{Spectral graphs}

Spectral graph theory studies the spectra of  graph Laplacian matrices. It  gives rise to the topological and spectral properties of underlying graphs or networks. Mathematically, a graph is an ordered pair $G(V,E)$, where $V=\{v_i; i=1,2,...,N\}$ is the vertex set with size $N$ and $E=\{e_{ij}=(v_i,v_j);i\le i<j\le N\}$ is the edge set. Denote $\text{deg}(v)$ the degree of each vertex $v_i\in V$, i.e.,  the number of edges that connects to $v$. A specific   Laplacian matrix $L^G$ can be given by
\begin{equation}
L^G = 
\begin{cases}
\text{deg}(v),& \text{if } v_i=v_j,\\
-1,           & \text{if } v_i \text{ and } v_j \text{ are adjacent},\\
0,            & \text{otherwise}, 
\end{cases}
\end{equation}
where ``adjacent'' is subject to a specific definition or connection rule. 

Let order the eigenvalues of the graph Laplacian matrix  as
\begin{equation}
\lambda_\text{min}=\lambda_1\le\lambda_2\le\cdots\le\lambda_N=\lambda_\text{max}.
\end{equation}
The kernel dimension of $L^G$ is the multiplicity of 0 eigenvalues, indicating the number of connected components of $G(V,E)$, which is the topological property of the graph. The non-zero eigenvalues of $L^G$    contain  the graph properties. 
In particular, $\lambda_2$ is called the algebraic connectivity.   

\subsubsection{Simplicial complex}

To construct a topological description of a graph, simplicial complex is used.   For a set of $q+1$ points, $\{v_0, v_1, ..., v_q\}$, a $q$-plane is well defined if the $q+1$ points are affinely independent, i.e., $v_1-v_0$, $v_2-v_0$, $...$, $v_q-v_0$ are linearly independent. Thus, one can have at most $n$ linearly independent vectors with at most $n+1$ affinely independent points in $\mathbb{R}^n$. An affine hull is the set of affine combinations, $v=\sum_{i=0}^{q}c_iv_i$, $c_i\in\mathbb{R}$, and $\sum_{i=0}^{q}c_i=1$. Such an affine combination is a convex combination if all $c_i$ are non-negative. The convex hull is the set of convex combinations. A $q$-simplex denoted as $\sigma_q$ is the convex hull of $q+1$ affinely independent points. For example, $0$-, $1$-, $2$-, and $3$-simplex are vertexes, edges, triangles, and tetrahedrons. A simplicial complex $K$ is a collection of simplices in $\mathbb{R}^n$ satisfying the following conditions such as the Cech complexes, Vietoris-Rips complexes, and alpha shapes. For example, the Vietoris-Rips complex of $K$ with radius $r$ consists of all subsets of radius $\text{R}(\sigma)$ at most $r$ as
\begin{equation}
\text{VR}(r) = \{\sigma \subseteq K | \text{R}(\sigma) \le r\}.
\end{equation} 
For $\sigma_q\in K$, its face $\sigma_{q-1}$ is also in $K$. The non-empty intersection of any two simplices $\sigma_q$, $\sigma_p\in K$ is a face of them. The dimension of simplicial complex is defined as the maximum dimension of its simplex. 

A $q$-chain is a finite sum of simplices as $\Sigma_{i}c_i\sigma_i^k$ with $\mathbb{Z}_2$ field of the coefficients $c_i$ for the sum, and the set of all chains in a group $C_q(K)$. The boundary operator $\partial_k$ maps $C_q(K)\rightarrow C_{q-1}(K)$ defined as
\begin{equation}
\partial_q\sigma_q=\sum_{i=0}^{q}(-1)^i[v_0, ..., \hat{v}_i,...,v_k]=\sum_{i=0}^{q}(-1)^i\sigma_{q-1}^i,
\end{equation}
where $\sigma_q=[v_0, {v}_1,...,v_k]$ and $\hat{v}_i$ stands for $v_i$ being omitted. A $q$-chain is called $q$-cycle if its boundary is zero. A chain complex is the sequence of chain groups connected by boundary operators
\begin{equation}
\cdots \stackrel{\partial_{q+2}}\longrightarrow C_{q+1}(K) \stackrel{\partial_{q+1}}\longrightarrow C_{q}(K) \stackrel{\partial_{q}}\longrightarrow C_{q-1}(K)\stackrel{\partial_{q-1}} \longrightarrow \cdots
\end{equation}
and the $k$-th homology group $H_k$ is defined by $H_k = Z_k / B_k$ where $Z_k={\rm ker} ~\partial_k=\{c\in C_k \mid \partial_k c=0\}$ and $B_k={\rm im} ~\partial_{k+1}= \{ \partial_{k+1} c \mid c\in C_{k+1}\}$. The Betti numbers are defined by the ranks of $k$-th homology group $H_k$. This, in practice, is counting holes in $k$-dimension, such as $\beta_0$ reflects the number of connected components, $\beta_1$ gives the number of loops, and $\beta_2$ is the number of cavities. In a nutshell, the Betti number sequence $\{\beta_0, \beta_1, \beta_2, \cdots\}$ reveals the intrinsic topological property of the system. 

Recall that in graph theory, the degree of a vertex ($0$-simplex) $v$ is the number of edges that are adjacent to the vertex, denoted as deg$(v)$. However, once we generalize this notion to $q$-simplex, problem aroused since $q$-simplex can have $(q-1)$-simplices and  $(q+1)$-simplices adjacent to it at the same time. Therefore, the upper adjacency  and lower adjacency are required to define the degree of a $q$-simplex for $q>0$  \cite{serrano2019centrality,Maletic2014}.
\begin{Def}
	Given two $q$-simplices $\sigma^i_{q}$ and $\sigma^j_{q}$ of a simplicial complex $K$. We say they are lower adjacent if they share a common $(q-1)$-face, denoted as $\sigma^i_q \stackrel{L}\sim \sigma^j_q$. The lower degree of $q$-simplex is the number of nonempty $(q-1)$-simplices in $K$ that are faces of $\sigma_q$, which is denoted as deg$_L(\sigma_q)$ and is always $q+1$.
\end{Def}
\begin{Def}
	Given two $q$-simplices $\sigma^i_{q}$ and $\sigma^j_{q}$ of a simplicial complex $K$. We say they are upper adjacent if they share a common $(q+1)$-face, denoted as $\sigma^i_q \stackrel{U}\sim \sigma^j_q$. The upper degree of $q$-simplex is the number of $(q+1)$-simplices in $K$ of which $\sigma_q$ is a face, which is denoted deg$_U(\sigma_q)$.
\end{Def}
Then, the degree of a $q$-simplex ($q>0$) is defined as:
\begin{equation}
\text{deg}(\sigma_q) = \text{deg}_L(\sigma_q) + \text{deg}_U(\sigma_q) = \text{deg}_U(\sigma_q) + q + 1.
\end{equation}

\subsubsection{Graph Laplacian}
The graph Lapalcian was introduced to enrich topological and geometric information of simplicial complexes via a filtration process. The preliminary concepts are about the oriented simplicial complex and $q$-combinatorial Laplacian. More detail information can be found elsewhere\cite{hernandez2019higher, Maletic2014, goldberg2002combinatorial, horak2013spectra}. The properties of the $q$-combinatorial Laplacian matrix with its spectra are discussed in the following.

A $q$-combinatorial Laplacian is defined based on oriented simplicial complexes, and its lower- and higher-dimensional simplexes can be employed to study a specifically oriented simplicial complex. An oriented simplicial complex $K$ is defined if all of its simplices are oriented. If $\sigma_q^i$ and $\sigma_q^j$ are upper adjacent with a common upper $(q+1)$-simplex $\tau_{q+1}$, they are similarly oriented if both have the same sign in  $\partial_{q+1}(\tau_{q+1})$ and dissimilarly oriented if the signs are opposite. Additionally, if $\sigma_q^i$ and $\sigma_q^j$ are lower adjacent with a common lower $(q-1)$-simplex $\eta_{q-1}$, they are similarly oriented if $\eta_{q-1}$ has the same sign in  $\partial_q(\sigma_q^i)$ and $\partial_q(\sigma_q^j)$, and dissimilarly oriented if the signs are opposite. Similarly, $q$-chains can be defined on the oriented simplicial complex $K$, as well as $q$-boundary operator. 

The $q$-combinatorial Laplacian is a linear operator $\Delta_q: C_q(K) \longrightarrow C_q(K)$ for integer $q\ge 0$
\begin{equation}\label{equ:laplacian operator}
\Delta_q := \partial_{q+1} \partial_{q+1}^{\ast} + \partial_{q}^{\ast} \partial_{q}
\end{equation}
where $\partial_{q}^{\ast}$ is the coboundary operator mapping $\partial_q^{\ast}: C_{q-1}(K) \longrightarrow C_q(K)$.
One property $\partial_q \partial_{q+1} = 0$ is preserved, which implies $\text{Im}(\partial_{q+1}) \subset \text{ker}(\partial_q)$.
The $q$-combinatorial Laplacian matrix, denoted $\mathcal{L}_q$, is the matrix representation.
\begin{equation}\label{equ:combinatorial Laplacian}
\mathcal{L}_q = \mathcal{B}_{q+1}\mathcal{B}_{q+1}^{T} + \mathcal{B}_q^T \mathcal{B}_q
\end{equation}
of operator $\Delta_q$, where $\mathcal{B}_q$ and $\mathcal{B}_q^T$ be the matrix representation of a $q$-boundary operator and $q$-coboundary operator, respectively, with respect to the standard basis for $C_q(K)$ and $C_{q-1}(K)$ with some assigned orderings. Then, the number of rows in $\mathcal{B}_q$ corresponds to the number of $(q-1)$-simplices and the number of columns shows the number of $q$-simplices in $K$, respectively. 
In addition, the upper and lower $q$-combinatorial Laplacian matrices are denoted by $\mathcal{L}_q^U = \mathcal{B}_{q+1}\mathcal{B}_{q+1}^{T}$ and $\mathcal{L}_q^L = \mathcal{B}_q^T \mathcal{B}_q$, respectively. Note that $\partial_0$ is the zero map which leads to $\mathcal{B}_0$ being a zero matrix. Therefore, $\mathcal{L}_0(K) = \mathcal{B}_{1} \mathcal{B}_1^T + \mathcal{B}_0^T \mathcal{B}_0$, with $K$ the (oriented) simplicial complex of dimension $1$, which is actually a simple graph. Especially, $0$-combinatorial Laplacian matrix $\mathcal{L}_0(K)$ is actually the Laplacian matrix defined in the spectral graph theory.

Given an oriented simplicial complex $K$ with $0 \le q \le \text{dim}(K)$, 
the entries of $q$-combinatorial Laplacian matrices are given by \cite{goldberg2002combinatorial}
\begin{align}
q>0,\ (\mathcal{L}_q)_{ij} &=
\begin{cases}\label{equ:combine}
\text{deg}(\sigma^i_{q}), & \mbox{if $i=j$.} \\
1,                 & \mbox{if $i\neq j$, $\sigma^i_q \stackrel{U}\nsim \sigma^j_q$ and $\sigma^i_q \stackrel{L}\sim \sigma^j_q$ with similar orientation.}   \\
-1,                & \mbox{if $i\neq j$, $\sigma^i_q \stackrel{U}\nsim \sigma^j_q$ and $\sigma^i_q \stackrel{L}\sim \sigma^j_q$ with dissimilar orientation.} \\
0,                 & \mbox{if $i\neq j$ and either , $\sigma^i_q \stackrel{U}\sim \sigma^j_q$ or $\sigma^i_q \stackrel{L}\nsim \sigma^j_q$.}
\end{cases} \\
q=0,\ (\mathcal{L}_q)_{ij}  &=
\begin{cases}\label{equ:L0}
\text{deg}(\sigma^i_{0}),   & \mbox{if $i=j$.} \\
-1,                         & \mbox{if $\sigma^i_0 \stackrel{U}\sim \sigma^j_0$.} \\
0,                          & \mbox{otherwise.}
\end{cases}
\end{align}

\subsubsection{Persistent spectral graphs}
Persistent spectral graphs were introduced by integrating graph Laplacian and multiscale filtration \cite{wang2020persistent}. Both topological and geometric information (i.e. connectivity and robustness of simple graphs) can be derived from analyzing the spectra of $q$-combinatorial Laplacian matrix. However, this method is genuinely free of metrics or coordinates, which induced too little topological and geometric information that can be used to describe a single configuration. Therefore, persistent spectral graphs (PSG) is proposed to create a sequence of simplicial complexes induced by varying a filtration parameter, which is inspired by the idea of persistent homology and our earlier work in multiscale graphs. This section mainly introduce the construction of persistent spectral graphs.

First, a $q$-combinatorial Laplacian matrix is symmetric and positive semi-definite. Therefore, its eigenvalues are all real and non-negative. The multiplicity of zero spectra (also called harmonic spectra) reveals the topological information, and the geometric information will be preserved in the non-harmonic spectra. More specifically, the multiplicity of zero spectra of $\mathcal{L}_q(K)$ is denoted by $\beta_q$ which is actually the $q$-th Betti number defined in the homology:
\begin{equation}\label{equ:betti}
\beta_q = \text{dim}(\mathcal{L}_q(K)) - \text{rank}(\mathcal{L}_q(K)) = \text{nullity}(\mathcal{L}_q(K)) = \# ~{\rm of ~ zero ~eigenvalues~ of}~   \mathcal{L}_q(K).
\end{equation}

Naturally, persistent spectral theory creates a sequence of simplicial complexes induced by varying a filtration parameter \cite{wang2020persistent}. A filtration of an oriented simplicial complex $K$ is a sequence of sub-complexes $(K_t)_{t=0}^m$ of $K$
\begin{equation}
\emptyset = K_0 \subseteq K_1 \subseteq K_2 \subseteq \cdots \subseteq K_m = K.
\end{equation}
It induces a sequence of  chain complexes
\begin{equation}
\left.\begin{array}{cccccccccccccc}
\cdots & C_{q+1}^1 &
\xrightleftharpoons[\partial_{q+1}^{1^\ast}]{\partial_{q+1}^1} & C_q^1 &
\xrightleftharpoons[\partial_q^{1^\ast}]{\partial_q^1} & \cdots & \xrightleftharpoons[\partial_3^{1^\ast}]{\partial_3^1} & C_2^1 & \xrightleftharpoons[\partial_2^{1^\ast}]{\partial_2^1} & C_1^1 & \xrightleftharpoons[\partial_1^{1^\ast}]{\partial_1^1} & C_0^1 & \xrightleftharpoons[\partial_0^{1^\ast}]{\partial_0^1} & C_{-1}^1 \\
& \rotatebox{-90}{$\subseteq$} &  & \rotatebox{-90}{$\subseteq$} &  &  &  & \rotatebox{-90}{$\subseteq$} &  & \rotatebox{-90}{$\subseteq$} &  & \rotatebox{-90}{$\subseteq$} &  &  \\
\cdots & C_{q+1}^2 &
\xrightleftharpoons[\partial_{q+1}^{2^\ast}]{\partial_{q+1}^2} & C_q^2 &
\xrightleftharpoons[\partial_q^{2^\ast}]{\partial_q^2} & \cdots &
\xrightleftharpoons[\partial_3^{2^\ast}]{\partial_3^2} & C_2^2 & \xrightleftharpoons[\partial_2^{2^\ast}]{\partial_2^2} & C_1^2 & \xrightleftharpoons[\partial_1^{2^\ast}]{\partial_1^2} & C_0^2 & \xrightleftharpoons[\partial_0^{2^\ast}]{\partial_0^2} & C_{-1}^1 \\
& \rotatebox{-90}{$\subseteq$} &  & \rotatebox{-90}{$\subseteq$} &  &  &  & \rotatebox{-90}{$\subseteq$} &  & \rotatebox{-90}{$
	\subseteq$} &  & \rotatebox{-90}{$\subseteq$} &  &  \\
& \rotatebox{-90}{$\cdots$} &  & \rotatebox{-90}{$\cdots$} &  &  &  & \rotatebox{-90}{$\cdots$} &  & \rotatebox{-90}{$\cdots$} &  & \rotatebox{-90}{$\cdots$} &  &  \\
& \rotatebox{-90}{$\subseteq$} &  & \rotatebox{-90}{$\subseteq$} &  &  &  & \rotatebox{-90}{$\subseteq$} &  & \rotatebox{-90}{$
	\subseteq$} &  & \rotatebox{-90}{$\subseteq$} &  &  \\
\cdots & C_{q+1}^m &
\xrightleftharpoons[\partial_{q+1}^{m^\ast}]{\partial_{q+1}^m} & C_q^m &
\xrightleftharpoons[\partial_q^{m^\ast}]{\partial_q^m} & \cdots &
\xrightleftharpoons[\partial_3^{m^\ast}]{\partial_3^m} & C_2^m & \xrightleftharpoons[\partial_2^{m^\ast}]{\partial_2^m} & C_1^m & \xrightleftharpoons[\partial_1^{m^\ast}]{\partial_1^m} & C_0^m & \xrightleftharpoons[\partial_0^{m^\ast}]{\partial_0^m} & C_{-1}^1
\end{array}\right.
\end{equation}

For each sub-complexes $K_t$, we define its corresponding chain group to be $C_q(K_t)$, and the $q$-boundary operator will be denoted by $\partial_q^t: C_q(K_t) \to C_{q-1}(K_t)$. We say that if $q<0$. then $C_q(K_t)$ is an empty set and $\partial_q^t$ is a zero map. If $0<q \le \text{dim}(K_t)$, then
\begin{equation}
    \partial_q^t(\sigma_q) = \sum_{i}^q(-1)^i \sigma^i_{q-1}, \sigma_q \in K_t,
\end{equation}
with $\sigma_q = [v_0, \cdots, v_q]$ being the $q$-simplex, and $\sigma^{i}_{q-1} = [v_0, \cdots, \hat{v_i} ,\cdots,v_q]$ being the $(q-1)$-simplex for which its vertex $v_i$ is removed. Additionally, the adjoint operator is $\partial_q^{t^{\ast}}: C_{q-1}(K_t) \to C_q(K_t)$. The topological and spectral information of $K_t$ can be analyzed from $\mathcal{L}_q(K_t)$ along with the filtration parameter by diagonalizing the $q$-combinatorial Laplacian matrix. We call the multiplicity of zero spectra of $\mathcal{L}_q^{t}$ as its persistent Betti number $\beta_q^{t}$, which counts the number of $q$-dimensional holes in $K_t$:
\begin{equation}
    \beta_q^{t} = \text{dim}(\mathcal{L}_q^{t}) - \text{rank}(\mathcal{L}_q^{t}) = \text{nullity}(\mathcal{L}_q^{t}) = \# \text{of harmonic spectra of } \mathcal{L}_q^{t}.
\end{equation}
Specifically, $\beta_0^t$ represents the number of connected components in $K_t$, $\beta_1^t$ reveals the number of one-dimensional loops or circles in $K_t$, and $\beta_2^t$ shows the number of two-dimensional voids or cavities in $K_t$. Moreover, the set of spectra of $\mathcal{L}_q^{t}$ is given by:
\begin{equation}
\text{Spectra}(\mathcal{L}_q^{t}) = \{(\lambda_1)_q^{t}, (\lambda_2)_q^{t}, \cdots, (\lambda_N)_q^{t}  \},
\label{eqn:Spec}
\end{equation}
where $\mathcal{L}_q^{t}$ has dimension $N\times N$ and spectra are arranged in ascending order. The smallest non-zero eigenvalue of $\mathcal{L}_q^{t}$ is defined as $(\tilde{\lambda}_2)_q^{t}$. The $p$-persistent $q$-combinatorial Laplacian operator is defined by extending the boundary operator. Detailed descriptions can be found in Ref. \cite{wang2020persistent}.

\subsection{Predictive models for mutation-induced protein-protein binding free energy changes}

Since the harmonic spectra produced by the kernel of a persistent Laplacian contain exact topological information as that of persistent homology. As such, we utilize a persistent homology software, GUDHI, to generate purely topological representations of PPIs in dimensions 0, 1, and 2. Additionally, persistent Laplacian spectra, including both harmonic and non-harmonic parts, are  coded in Python. Machine learning and deep learning algorithms are implemented in Pytorch \cite{paszke2019pytorch}.

\subsubsection{Persistent Laplacian representation of PPIs}
 
To facilitate   topological  and shape analysis of PPIs via persistent Laplacians, we first composite the atoms in a protein-protein complex into various subsets. 
\begin{enumerate}
	\item $\mathcal{A}_\text{m}$: atoms of the mutation sites.
	\item $\mathcal{A}_\text{mn}(r)$: atoms in the neighbourhood of the mutation site within a cut-off distance $r$.
	\item $\mathcal{A}_\text{A}(r)$: protein A atoms within $r$ of the binding site.
	\item $\mathcal{A}_\text{B}(r)$: protein B atoms within $r$ of the binding site.
	\item $\mathcal{A}_\text{ele}(\text{E})$: atoms in the system that has atoms of element type E. The distance matrix is specially designed such that it excludes the interactions between the atoms form the same set. For interactions between atoms $a_i$ and $a_j$ in set $\mathcal{A}$ and/or set $\mathcal{B}$, the modified distance is defined as
	\begin{equation}
	D_{\text{mod}}(a_i, a_j) =
	\begin{cases}
	\infty, \text{ if } a_i, a_j\in\mathcal{A}\text{, or }a_i, a_j\in \mathcal{B}, \\
	D_e(a_i, a_j), \text{ if } a_i\in\mathcal{A} \text{ and }a_j\in \mathcal{B},
	\end{cases}
	\label{eq:modified_equation}
	\end{equation}
	where $D_e(a_i, a_j)$ is the Euclidian distance between $a_i$ and $a_j$.
\end{enumerate}
Molecular atoms of different can be constructed as points presented by $v_0$, $v_1$, $v_2$, $...$, $v_k$ as $k\!+\!1$ affinely independent points in simplicial complex. 
Persistent spectral graph is devised to track the multiscale topological and geometrical information over different scales along a filtration \cite{wang2020persistent}, resulting  in significant important feature vectors for the machine learning method. 
Features generated by binned barcode vectorization can reflect the strength of atom bonds, van der Waals interactions, and can be easily incorporated into a machine learning model, which captures and discriminates local patterns. 
Using the atom subsets, for example $\mathcal{A}_\text{A}(r)$ and $\mathcal{A}_\text{B}(r)$, simplicial complexes are constructed by only considering the edges from $\mathcal{A}_\text{A}(r)$ to $\mathcal{A}_\text{B}(r)$ for Vietoris-Rips complexes. Then from the Vietoris-Rips complex filtration, barcodes generated from persistent homology are enumerated by bar lengths in certain intervals with number 0 or 1. Meanwhile, for each complexes in the filtration, eigenvalues are calculated according to the graph Laplacian analysis. The statistics of eigenvalues such as sum, maximum, minimum, mean, and standard deviation are collected to have a normalized features for machine learning methods.
Another method of vectorization is to get the statistics of bar lengths, birth values, and death values, such as sum, maximum, minimum, mean, and standard deviation. 
This method is applied to vectorize Betti-1 ($H_1$) and Betti-2 ($H_2$) barcodes obtained from alpha complex filtration based on the facts that higher-dimensional barcodes are sparser than $H_0$ barcodes.

\subsubsection{Machine learning and deep learning algorithms}

The features generated from the persistent spectral graph are tested by the gradient boosting tree (GBT) method and the deep neural network (Net) method. The validations are performed on the datasets discussed in the results section. The accurate prediction of the mutation-induced binding affinity changes  of protein-protein complexes is very challenging. After effective feature-generations, a  machine learning or deep learning model is also required for validations and real applications. The gradient boosting tree is a popular  ensemble method  for regression and classification problems. It builds a sequence of  weak learners to correct training errors. By the assumption that the individual learners are likely to make different mistakes, the method combines weak learners to eliminate the overall error. Furthermore, a decision tree is added to the ensemble depending on the present prediction error on the training dataset. Therefore, this method is relatively robust against hyperparameter tuning and overfitting, especially for a moderate number of features.  The GBT is shown for its robustness against overfitting, good performance for moderately small data sizes, and model interpretability. The present work uses the package provided by scikit-learn (v 0.23.0) \cite{pedregosa2011scikit}.
The number of estimators and the learning rate are optimized for ensemble methods as 20000 and 0.01, respectively. For each set, ten runs (with different random seeds) were done, and the average result is reported in this work. Considering a large number of features, the maximum number of features to consider is set to the square root of the given descriptor length for GBT methods to accelerate the training process. The parameter setting shows that the performance of the average of sufficient runs is decent. 

A deep neural network   is a network of neurons that maps an input feature layer to an output layer. The neural network mimics the human brain to solve problems with numerous neuron units with  backpropagation to update weights on each layer. To reveal the facts of input features at different levels and abstract more properties, one can construct more layers and more neurons in each layer, known as a deep neural network. Optimization methods for feedforward neural networks and dropout methods are applied to prevent overfitting. The network layers and the number of neurons in each layer are determined by gird searches based on 10-fold cross-validations. Then, the hyperparameters of stochastic gradient descent (SGD) with momentum are set up based on the network structure. The network has 7 layers with 10000 neurons in each layer. For SGD with momentum, the hyperparameters are \texttt{momentum = 0.9} and \texttt{weight\_decay=0}. The learning rate is 0.002 and the epoch is 400. The Net is implemented on Pytorch \cite{paszke2019pytorch}.

\subsubsection{Predictive models}

In our previous work,  topology-based deep neural network trained with mAbs (TopNetmAb) was introduced to predict mAb binding free energy changes \cite{chen2021revealing}. Persistent homology is the main workhorse for TopNetmAb, but auxiliary features inherited from our earlier TopNetTree \cite{wang2020topology} are utilized. 

In this work, we construct a TopNet model from TopNetmAb by excluding mAb training data. A topology-based GBT model (TopBGT) is also developed in the present work by replacing Net in the TopNet model with GBT.  Both TopNet and TopGBT include a set of auxiliary features inherited from our earlier TopNetTree \cite{wang2020topology} and TopNetmAb \cite{chen2021revealing} to enhance their performance. 

Additionally, to evaluate the performance of persistent Laplacian (Lap) for PPIs, we construct  persistent Laplacian-based GBT (LapGBT) and persistent  Laplacian-based deep neural network (LapNet). Note that unlike  TopNet and TopGBT,  LapGBT and LapNet employ only persistent  Laplacian features extracted from protein structures. Therefore, their performance depends purely on persistent Laplacian.  

Moreover, TopLapGBT and TopLapNet are constructed  by adding persistent Laplacian features to TopGBT and TopNet, respectively. Furthermore, the consensus of GBT and Net predictions are also used for validations, denoted  as TopNetGBT and LapNetGBT, respectively. Finally, the consensus of TopLapNet and  TopLapGBT is called TopLapNetGBT.

\section{Conclusion}

Due to natural selection, emerging SARS-CoV-2 variants are spreading worldwide with their increased transmissibility as a result of higher infectivity and/or  stronger antibody resistance. The increase in antibody resistance also leads to vaccine breakthrough infections and  jeopardizes the existing monoclonal antibody drugs.  The spike protein plays  the most important role in viral transmission because its receptor binding domain (RBD) binds to human ACE2 to facilitate the viral entry of  host cells. Topological data analysis (TDA) of RBD-ACE2 binding free energy changes induced by RBD mutations enables the accurate forecasting of emerging SARS-CoV-2 variants \cite{chen2020mutations, wang2021emerging, chen2022omicron, chen2022omicron2}. 

However, the earlier TDA method is not sensitive to homotopic shape evolution, which is important for protein-protein interactions (PPIs). To overcome this obstacle, persistent Laplacian, which characterizes the topology and shape of data, is introduced in this work for analyzing PPIs. Paired with advanced machine learning and deep learning algorithms, the proposed persistent Laplacian method outperforms  the state-of-art approaches in  validation with mutation-induced binding free energy changes of PPIs using major benchmark datasets. An important forecasting from the present work is that Omicron subvariants BA.2.11,  BA.212.1, BA.4, and BA.5 have  high potential to become new  dominating variants in the world. 

\section*{Acknowledgment}
This work was supported in part by NIH grant  GM126189, NSF grants DMS-2052983,  DMS-1761320, and IIS-1900473,  NASA grant 80NSSC21M0023,  Michigan Economic Development Corporation, MSU Foundation,  Bristol-Myers Squibb 65109, and Pfizer.

\pagebreak

\end{document}